\documentclass[onecolumn]{article}
\usepackage{graphicx}
\usepackage{marvosym}
\usepackage{amsmath,amsthm,amssymb}
\newtheorem{thm}{Theorem}[section]
\newtheorem{proposition}[thm]{Proposition}

\usepackage{array}
\usepackage{setspace}
\usepackage{amssymb}
\usepackage{amsthm}
\usepackage{epstopdf}
\usepackage[top=1in, bottom=1.5in, left=1in, right=1in]{geometry}
\usepackage{caption}
\usepackage{subcaption}

\begin{document}
\title{\bf{An original Propagator for large array}}
\author{Youssef Khmou, Said Safi.\\
Department of Mathematics and Informatics, Polydisciplinary faculty,\\
                   Sultan Moulay Slimane University,\\
                             Beni Mellal, Morocco.}
\maketitle

\begin{abstract}
In this paper, we demonstrate that when the ratio $n$ of the number of antenna elements $N$ to the number $P$ of radiating sources is superior or equal to 2, then it is possible to choose a propagator from a set of ${n(n+1)/2}-1$ operators to compute the Angles of Arrival (AoA) of the narrowband incoming waves.\\
This new non eigenbased approach is efficient when the Signal to Noise Ratio (SNR) is moderate, and gives multitude of possibilities, that are dependent of the random data, to construct the complex sets whose columns are orthogonal to the signal subspace generated by the radiating sources. Elementary examples are given for $n=3$, $n=4$ and $n=6$. The simulation results are presented to illustrate the performance of the proposed computational methods.
\end{abstract}
\section{Introduction}
\label{}
The new technology of planar arrays and nanoarrays consists of constructing thousands of radiating elements connected to single unit for processing the emitted and the received signals. With this new conception, the computational complexity [1] is proportional to the increasing rate of the antenna elements. When dealing with the problem of Angles of Arrival (AoA) detection [1], computing the second order statistics is sufficient, however the eigen-approach is computationally heavy. Recent studies focused on the asymptotic behavior of the eigenvalues of the cross correlation matrix when its dimensions tend to infinity [2], the distribution of the eigenvalues gives the repartition that can be clustered to obtain complex signal and noise subspaces, with the latter being characterized by spherical covariance matrix [1].\\

Besides the eigen-approach, other methods [3] were derived by exploiting the evolution of the wavefield along the antenna sensors, a linear operator relates the channel matrix of the first antenna elements to the rest of them, this operator is called Propagator [4] and can be extracted from the inter spectral matrix. Since the appearence of the propagator, few versions were proposed to generalize the concept, in [5] the channel matrix was descritized into several partitions where the partial propagator has non redundent elements while all the sensors are exploited. in [6], the propagator was extended into third order when the number of sensors is superior to two times the number of sources.\\

In this paper, we propose a generalization of the propagator for large antenna arrays, we demonstrate that given the number of sensor $N$ and the number of emitting sources $P$, if the ratio $n=E(N/P)$ is superior or equal to $2$, then it is possible to generate $n$ operators of the noise subspace and the sum of the computational methods is given by ${n(n+1)/2}-1$ possibilities. In the next section we present the statistical data model of the problem, in third section we develop the extended propagator and in the last section we present some numerical results.
\section{STATISTICAL DATA MODEL}
In this section we elaborate the physical description of the model used in this paper. An array antenna, consisting of $N$ identical and isotropic sensors, is receiving a wavefield that is generated by $P$ sources such that theirs frequency spectra are concentrated around a carrier frequency $f_{c}$. We assume that the velocities of the sources are negligible during the acquisition time $T$ of the array, $dx_{i}/dt \simeq 0$ for $i=1,...,P$, the sources are located in the far field region of the antenna. The medium of propagation has the same permettivity and permeability $(\epsilon,\mu)$ in all directions which permits the wavefronts of the sources to be summed linearly. If we consider that the noise in the medium, the electronic and thermal noises of the sensors are modeled by ergodic random processes, the received signals at instant 
$t_{k}$ can be written in the following form :
\begin{equation}
X(t_{k})=\sum_{j=1}^{N}\sum_{i=1}^{P}a_{j}(\theta_{i})s_{i}(t_{k})+n(t_{k})
\end{equation}
with $k=1,2,...,K$ and $K$ being the number of samples, $a(\theta_{i}) \in \mathbb{C}^{N\times 1 }$ is the steering vector of the $i^{th}$ angle of arrival, the progressive phase along the sensors is given by :
\begin{equation}
a(\theta_{i})={\left[ 1,e^{-j\mu_{i}}, e^{-j2\mu_{i}},..., e^{-j(N-1)\mu_{i}} \right] }^{T} 
\end{equation}
$(.)^{T}$ is the transposition operator and  $\mu_{i}=2\pi d \lambda^{-1} \sin{\theta_{i}}$, $d$ is the inter-element distance of the array which we consider to be uniform and equal to the half of the wavelength $\lambda$. $s_{i}(t)$  is the carried signal of the $i^{th}$ source with $s(t)={\left[s_{1}(t),s_{2}(t),...,s_{P}(t)\right]}^{T}$ and $n(t)={\left[n_{1}(t),n_{2}(t),...,n_{N}(t)  \right] }^{T}$ is the additive noise modeled by stationary and ergodic zero mean complex random process. In compact form, the received signals are given by :
\begin{equation}
X(t)=A(\theta)s(t)+n(t)
\end{equation}
$A(\theta)$ is the channel matrix of the array. Based on the matrix $X(t)$ we can extract a lot of characteristics of the studied system, such as the waveforms of the signals $s(t)$, the polarization of the sources, the coherence between the signals $s(t)$ and the number of the signals $P$.\\
In this paper we are focused on detecting the Angles of Arrival (AoA) of the punctual emitters $\{\theta_{i}\}$, for this purpose, many methods can be applied to estimate the angles, almost all the techniques are based on second order statistics of $X(t)$ which can be computed by :
\begin{equation}
\Gamma=\underset{K\to +\infty}{\lim} {1\over K} \sum_{t=1}^{K} X(t) X^{+}(t)
\end{equation}
With finite number of samples $K$ only an approximation is obtained. The theoretical expression of $\Gamma$ is given by the following equation :
\begin{equation}
\Gamma=A(\theta)\Gamma_{s}A^{+}(\theta)+\Gamma_{n}
\end{equation}
the matrix $\Gamma_{s} \in \mathbb{C}^{P \times P}$ is the covariance of waveforms, $\Gamma_{n}=\sigma^{2} I_{N}$ is the covariance of the random vector $n(t)$, $(.)^{+}$ is the conjugate transpose operator, $\sigma^{2}$ is the noise variance considered to have the same value for all sensors and $I_{N}$ is the identity matrix of dimension $N$. The subspace based techniques [7-8] exploit the orthogonality between the vectorial spaces generated by the signal and noise eigenvectors, indeed the spectral theorem is applied to the covariance matrix, so that it can be decomposed into the following form :
\begin{equation}
\Gamma=\sum_{i=1}^{N}\lambda_{i}u_{i}u^{+}_{i}=U_{s}\Phi_{s} U_{s}^{+}+U_{n}\Phi_{n} U_{n}^{+}
\end{equation}
$U_{s} \in \mathbb{C}^{N\times P}$ is the vectorial space whose columns span the signal subspace and $U_{n} \in \mathbb{C}^{N\times N-P}$ is the set whose columns span the noise subspace. The diagonal matrices $\Phi_{s}$,$\Phi_{n}$ are given by $\Phi_{s}=diag\{\lambda_{1},...,\lambda_{P} \}$ which contains the largest eigenvalues in decreasing order $\lambda_{1} \geq \lambda_{2} \geq ...\geq \lambda_{P} $, $\Phi_{n}=diag\{\lambda_{P+1},...,\lambda_{N} \}$ where the diagonal elements are the noise eigenvalues, in the noiseless case we have $\lambda_{P+1}=...=\lambda_{N}=0$. The two orthogonal subspaces are related by the following equation :
\begin{equation}
U_{s}U^{+}_{s}+ U_{n}U^{+}_{n}=I_{N}
\end{equation}
In compact form, the spectral decomposition can be written as :
\begin{equation}
\Gamma=\left [\begin{array}{cc}
U_{s} & U_{n}
\end{array} \right ]
\left [ \begin{array}{cc}
\Phi_{s} &         \\
         & \Phi_{n}
\end{array} \right ]
{\left [\begin{array}{cc}
U_{s} & U_{n}
\end{array} \right ]}^{+}
\end{equation}
With the spectral decomposition, the degenerescence $d$ of the eigenvalues is determined in order to detect the number of sources $P=N-d$ by which the subspace $U_{n}$ is delimited to compute the angular spectrum [7] :
\begin{equation}
f(\theta_{i})=\{ a^{+}(\theta_{i}) U_{n} U^{+}_{n} a(\theta_{i})\}
\end{equation}
for testing angle $\theta_{i}$. If the number P is known, the subspace $U_{n}$ can be computed using the propagator function [4], this operator is described using the channel matrix, theoretically $A(\theta)$ can be divided into two blocks :
\begin{equation}
A= \left [ \begin{array}{c}
A_{1} \\
---  \\
A_{2}
\end{array} \right ]
\end{equation}
$A_{1} \in \mathbb{C}^{P\times P}$ and $A_{2} \in \mathbb{C}^{N-P\times P} $, given that the two blocks are linearly dependent, a linear operator $\Pi \in \mathbb{C}^{N-P\times P}$ exist such that $A_{2}=\Pi A_{1}$ and $\Pi=A_{2}A^{-1}_{1}$, consequently the vectorial space whose columns span the noise subspace $U_{n}$ is given by
\begin{equation}
\left [ \Pi \mid -I_{N-P} \right ] \left [A(\theta) \right ]=0_{N-P \times P}
\end{equation}
This operator is extracted from the covariance matrix, similarly  $\Gamma$ is split into two blocks $\Gamma =\left [ \Gamma_{s} \mid \Gamma_{n} \right ]$ with $\Gamma_{s} \in \mathbb{C}^{N\times P}$ , $\Gamma_{n} \in \mathbb{C}^{N\times N-P}$, an estimate of the propagator is given by the following equation :
\begin{equation}
\Pi^{+}=\Gamma^{\dagger}_{s} \Gamma_{n}
\end{equation}
with $(.)^{\dagger}$ is the Penrose pseudo inverse, $\Gamma^{\dagger}_{s}=(\Gamma^{+}_{s} \Gamma_{s})^{-1} \Gamma^{+}_{s}$. If we consider the noiseless case, using the equation (10), the covariance matrix is decomposed into four blocks :
\begin{equation}
\Gamma= \left [ \begin{array}{cc}
  \Gamma_{11}& \Gamma_{12} \\
\Gamma_{21}  &  \Gamma_{22}
\end{array}     \right ]
\end{equation}
with $\Gamma_{11} \in \mathbb{C}^{P\times P}$, $\Gamma_{21} \in \mathbb{C}^{N-P\times P}$, $\Gamma_{12} \in \mathbb{C}^{P\times N-P}$ and $\Gamma_{22} \in \mathbb{C}^{N-P\times N-P}$. Following the decomposition of the channel matrix [5], the $\Gamma$  blocks are computed by :
$\Gamma_{22}=A_{2}\Gamma_{s} A^{+}_{2}$, $\Gamma_{11}=A_{1}\Gamma_{s} A^{+}_{1}$ and $\Gamma_{21}=A_{2}\Gamma_{s} A^{+}_{1}$. We can obeserve that $\Gamma_{21}=\Pi \Gamma_{11}$ which leads to a noise subspace :
\begin{equation}
\left [ \Gamma_{21}\Gamma^{-1}_{11} \mid -I_{N-P} \right ] \left [ A(\theta) \right ] = 0_{N-P \times P}
\end{equation}

Usually in the litterature [4],[9-14] the propagator is developped using the equation (12), the second approach in equation (14) was proposed in [5], however we can deduce another version based on the equation $\Pi^{\dagger}={(A_{2}A^{-1}_{1})}^{\dagger}$, the corresponding subspace is described by the following equation :
\begin{equation}
\left [-I_{P} \mid \Gamma_{12}\Gamma^{\dagger}_{22}\right ] \left [ A(\theta) \right ] = 0_{P \times P}
\end{equation}
All the three versions are equivalent, but in the presence of the matrix $n(t)$ and the variation of the numbers $N$ and $P$, they are different due to the presence of the noise powers in the diagonal blocks $\Gamma_{11}$ and $\Gamma_{22}$. Given this situation, if we are supposed to work with a new architecture of the arrays comprising of hundreds of radiating elements which might be the case for nanoarrays, the question that we can ask is given a large number $N$ ( $N>>P$) can we find other possibilities for computing the propagator operator without using the diagonal blocks $\Gamma_{ii}$ ? the answer of this question is dependent on the ratio $n=N/P$ which will give a choice to choose an operator from a set that we will elaborate in the next section.
\section{Extended Propagator}
In this section, we develop our approach based on Uniform Linear Array (ULA) which is considered to be centro-symmetric, taking the first  element of the array as the origin of coordinates, the channel matrix has a Vandermonde structure.
\begin{equation}
A= \left ( \begin{array}{cccc}
1                 & 1                    & ... & 1\\
e^{-j\mu_{1}}     & e^{-j\mu_{2}}        & ... & e^{-j\mu_{P}}\\
...               &      ...             & ... & ...\\
e^{-j(N-1)\mu_{1}}& e^{-j(N-1)\mu_{2}}   &...  & e^{-j(N-1)\mu_{P}}
\end{array} \right )
\end{equation}
where $A$ has the properties $A \odot A^{*}=1_{N \times P}$, $rank(A)=P$ and $J_{N}A^{*}=A\Lambda_{P}$
where $J_{N}$ is the exchange matrix given by :
\begin{equation*}
J=\left ( \begin{array}{rrrrr}
0 & 0 & ... & 0 & 1\\
0 & 0 & ... & 1 & 0\\
... & ... & ... & ... & ...\\
0 & 1 & ... & 0 & 0\\
1 & 0 & ... & 0 & 0 
\end{array} \right )
\end{equation*}
$\Lambda_{P}$ is $P\times P$ diagonal matrix $\Lambda_{P}=diag\{e^{-j(N-1)\mu_{1}},...,e^{-j(N-1)\mu_{P}}\}$, we note that the rank of A is dependent on the azimuths of the sources $\theta_{i}$. Let us consider the three approaches for the propagator presented earlier:
\begin{equation}
\left\{ \begin{array}{rrr}
Q=\left [ {(\Gamma^{\dagger}_{s} \Gamma_{n})}^{+} \mid -I_{N-P} \right ]\\
Q_{1}=\left [ \Gamma_{21} \Gamma^{-1}_{11}\mid -I_{N-P} \right ]\\
Q_{2}=\left [ -I_{P} \mid\Gamma_{12} \Gamma^{\dagger}_{22}\right ] 
\end{array} \right.
\end{equation}
The maximum number of sources for $Q$ to resolve is $N-1$, which is also the case for $Q_{1}$, however if $N<2P$,  $Q_{2}$ will fail to compute the null space of $A(\theta)$, this problem is due to the density of matrix $Q_{2}$ which contains $P(P-1)$ zeros, if we compute the value of spectrum with arbitrary vector $a(\theta_{i})$ then the function $a(\theta_{i})Q_{2}^{+}Q_{2}a^{+}(\theta_{i})$ will not give us information whether $\theta_{i}$ is an Angle Of Incidence (AOI) or not.\\
Supposing that the Signal to Noise Ratio (SNR) is high, evaluating $Q_{2}$ requires the condition $N/P \geq 2$, following the same reasoning the ratio must also satisfy $n \leq \lfloor N/P \rfloor$, so the variable $n$ is bounded :
\begin{equation}
2\leq n \leq \lfloor{N\over P} \rfloor
\end{equation}
We introduce the selection matrices [16] to make the formalism rigourous, let us consider the set  $\{e_{i}\}_{1\leq i\leq n}$  where a matrix $e_{i} \in \mathbb{C}^{N\times P}$ is defined by $e_{i}={\left [0_{P \times P(i-1)}\mid I_{P}   \mid   0_{P\times N-(Pi)  } \right ] }^{T}$ which verifies the following equations :
\begin{equation}
e_{i}^{T}e_{j}= \delta_{ij} I_{P}
\end{equation}

\begin{equation}
\sum_{i=1}^{n} e_{i} e_{i}^{T} =I_{N}
\end{equation}

\begin{equation}
\sum_{i=1}^{n} e_{i}^{T}e_{i} di = I_{P}
\end{equation}

\begin{equation}
||e_{i} ||_{F} = \sqrt{P}
\end{equation}
where $di=1/n$ and $F$ denotes the Frobenius norm $(||e||_{F}=\sqrt{Tr(ee^{+})})$. The channel matrix is decomposed into :
\begin{equation}
A=\sum_{i=1}^{n} e_{i} e_{i}^{T}A=\sum_{i=1}^{n}e_{i}A_{i}
\end{equation}
This decomposition is equivalent to $A={\left [ A_{1}^{T}, A_{2}^{T},...,A_{n}^{T} \right ] }^{T}$,
the blocks $A_{i}$ of rank $P$  are equivalent of scalars in Euclidean base with the following property:
\begin{equation}
A_{i}=e_{i}^{T}A,\hspace{1cm} A^{+}_{i}=A^{+}e_{i}
\end{equation}
This partition is mapped into $n^{2}$ partitions on the covariance matrix $\Gamma$, this concept is the same principle used in the semiclassical approach of statistical mechanics where the one dimenisonal  space phase  is divided into cells on the condition  that the product of the step position and impulsion equals the quantum constant $h$, in our case the division of $\Gamma$ implies $dX dY=P^{2}$, this concept is illustrated in Fig.1.\\

\begin{figure}[!h]
\centering
\begin{subfigure}{6cm}
  \centering
  \includegraphics[width=6cm]{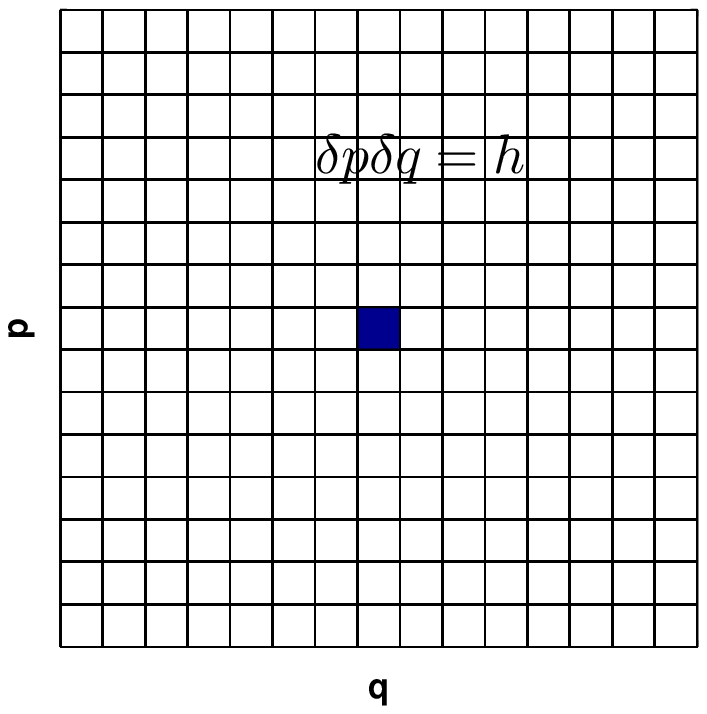}
  \caption{}
  \label{fig11}
\end{subfigure}%
\begin{subfigure}{6cm}
  \centering
  \includegraphics[width=6cm]{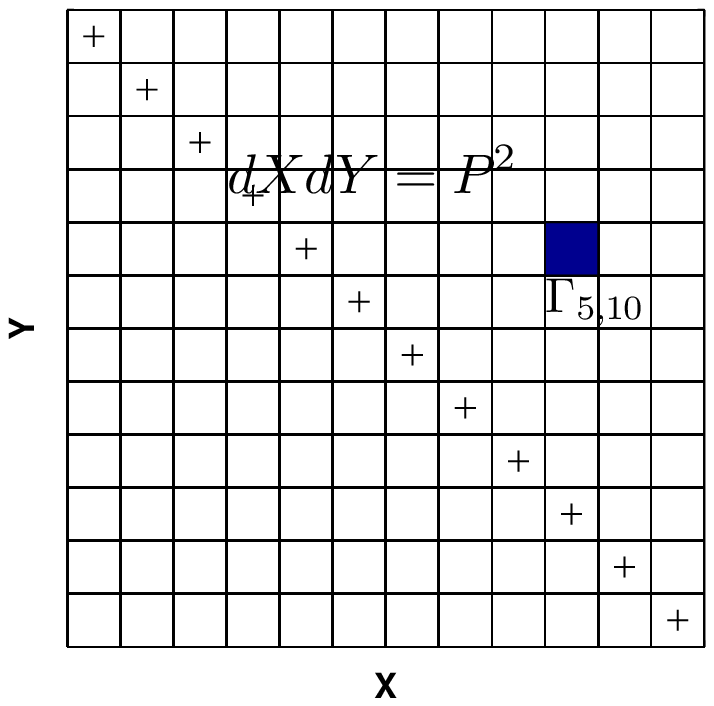}
  \caption{}
  \label{fig12}
\end{subfigure}
\caption{Similarity concept of partition principle between statistical physics and array processing : (a) semiclassical approach of one dimensional space phase $(p,q)$, (b) uniform partition of matrix $\Gamma$ where the mark $+$ denotes the affected blocks by the noise $\Gamma_{ii}$ and $N=12P$. }
\label{fig1}
\end{figure}

Given $n$ phases of $A$ there exist $n^{2}-n$ linear operators $Q_{ij}$ that relates the $i^{th}$ and $j^{th}$ phases $A_{i}=Q_{ij}A_{j}$. The reverse time evolution of the wavefronts  is introduced ($i<j$) unlike the standard propagator ($i=2>j=1$). The microstate between the $i^{th}$ and the $j^{th}$ sets of the sensors is given by :
\begin{equation}
\Gamma_{ij}=e_{i}^{T} \Gamma e_{j}=e_{i}^{T}A  \Gamma_{s} A^{+}e_{j}= A_{i}\Gamma_{s} A^{+}_{j}
\end{equation}

\begin{equation}
\tilde{\Gamma}_{ii}=e_{i}^{T}A  \Gamma_{s}A^{+}e_{i}+\sigma^{2} I_{P}
\end{equation}
Given the spectral matrix with the property $\Gamma^{+}=\Gamma$ the corresponding conjugate transpose block $\Gamma^{+}_{ij}$ is computed by :
\begin{equation}
\Gamma^{+}_{ij}={(e_{i}^{T} \Gamma e_{j})}^{+}=e_{j}^{T} \Gamma e_{i}=\Gamma_{ji}
\end{equation}
For infinite number of samples $K$, the general partition of $\Gamma$ is :
\begin{equation*}
\Gamma=\left ( \begin{array}{rrrrr}
\tilde{\Gamma}_{11}    & \Gamma_{12} & ... & ... & \Gamma_{1n}\\
\Gamma_{21} & \tilde{\Gamma}_{22}    & ... & ... & \Gamma_{2n}\\
...         &           ...                   & ... & ... & ...\\
...         &           ...                   & ... & ... & ...\\
\Gamma_{n1} &           ...                   & ... & ... & \tilde{\Gamma}_{nn}
\end{array} \right )
\end{equation*}
Where $\tilde{\Gamma}_{nn}=A_{n}\Gamma_{s}A^{+}_{n}+\sigma^{2}I_{N-(n-1)P}$, from equation (25), the partitions of the channel matrix are related by :
\begin{equation}
\left\{
\begin{array}{rr}
e_{i}^{T}A=\Gamma_{ik}\Gamma^{\dagger}_{jk}e_{j}^{T}A\\
A_{i}=\Gamma_{ik}\Gamma_{jk}^{\dagger}A_{j}
\end{array} \right.
\end{equation}
This equation proves that we have $(n-2)$ possibilities for arbitrary index $k \neq i \neq j$, $k$ can be chosen randomly because we assume the existence of isotropic noise $\Gamma_{n}=\sigma^{2} I_{N}$.
Using equation (28) we can derive the following equation :
\begin{equation}
\sum_{j=1,j\neq i}^{n}\Gamma_{ik} \Gamma^{\dagger}_{jk} e_{j}^{T}A-(n-1) e_{i}^{T}A=0_{P\times P}
\end{equation}
The above equation allows us to compute the $N\times P$ matrix whose columns span the null space of the channel matrix $A$ as follows :
\begin{equation}
\left \{{\sum_{j=1,j\neq 1}^{n} \Gamma_{ik} \Gamma_{jk}^{\dagger} e_{j}^{T}}-\beta_{P} e_{i}^{T} \right \} A=0_{P\times P}
\end{equation}
Where $\beta_{P}=(n-1) I_{P}$, the above equation is the $i^{th}$  propagator based on the $i^{th}$ block $A_{i}$, hence we have $n$ possibilities which can be concatenated in single matrix $\Psi \in \mathbb{C}^{N\times N}$ as follows :
\begin{equation}
\Psi = \left ( \begin{array}{ccccc}
-\beta_{P} & Q_{12}     & Q_{13} & ... & Q_{1n} \\
Q_{21}    & -\beta_{P}  & Q_{23} & ... & Q_{2n} \\
...       &  ...       & ...    & ... & ...\\
...       &  ...       & ...    & ... & ...\\
Q_{n1}    & Q_{n2}     & Q_{n3} & ... &  -\beta_{N-(n-1)P}
\end{array}     \right )
\end{equation} 
Given two blocks $(i,j)$ the propagator that maps the phase of the wavefront from block $i$ to the corresponding $j$ is product of partial propagators of blocks $i\leq k \leq j$ in both causal case $i>j$ and retarded case $i<j$:
\begin{equation}
Q_{ij}=\prod_{k=i}^{j-1} Q_{k,k-1}            \hspace{2cm}    Q_{ij}=\prod_{k=i}^{j-1} Q_{k,k+1}
\end{equation}
$\Psi$ has the property $Tr(\Psi)=-(n-1)N$ and each block $\Psi_{ij}=\Psi^{\dagger}_{ji}$ :
\begin{equation}
e_{i}^{T} \Psi e_{j}= {(e_{j}^{T} \Psi e_{i})}^{\dagger}
\end{equation}
To illustrate the general structure of $\Psi$, let us take an example of uniform linear array consisting of $N=3$ antennas and one radiating source $P=1$, the number of possibilities of partition is $n=\{2,3\}$, let us take $n=3$, the orthonormal base is reduced into $\{ e_{i}\}_{1\leq i\leq 3}$ where $\left [e_{1},e_{2},e_{3}\right ]=I_{3}$. The steering vector is given by 
\begin{equation}
A=1e_{1}+e^{j\phi_{1}}e_{2}+e^{2j\phi_{1}}e_{3}
\end{equation} 
with $\phi_{1}=2\pi \lambda^{-1}d \sin{\theta_{1}}$, applying the previous formalism yields to the following result of $\Psi$ :
\begin{equation}
\Psi= \left (  \begin{array}{ccc}
-2             & e^{-j\phi_{1}} &  e^{-2j\phi_{1}} \\
e^{j\phi_{1}}  &  -2            &  e^{-j\phi_{1}}  \\
e^{2j\phi_{1}} &  e^{j\phi_{1}} &  -2
\end{array}    \right )
\end{equation}
Hence the rows of $\Psi$ are orthogonal to the steering vector $A$ :
\begin{equation}
\Psi_{1}A=\Psi_{2}A=\Psi_{3}A=0
\end{equation}
We can deduce that $\Psi_{ij}=\Psi_{ji}^{-1}$ for single source $P=1$, $\Psi$ is self adjoint operator $\Psi^{+}=\Psi$.
The general rule elaborated is that for every integer $n$, there are $n^2$ partitions on $\Gamma$ and $n$ possible propagators are derived. Given sufficiently high SNR, we can state the following proposition :
\begin{proposition}
Given a random matrix $X \in \mathbb{C}^{N\times K}$ with $X \sim \mathcal{CN}(0_{N\times 1},\Gamma)$, $\exists n \in \mathbb{N}$ such that $\forall 2 \leq n \leq {\lfloor{N\over P} \rfloor}$, $\exists$  $\Omega=\left \{   \Psi_{i} \in \mathbb{C}^{N\times N}, i=1,...,n   \right \}$ with $Card(\Omega)=n(n+1)/2-1$ where we have :\\
\begin{equation}
\underset{SNR \to +\infty}{\lim}\Psi_{n}A=0_{N\times P}
\end{equation}
\begin{equation}
\underset{SNR \to +\infty}{\lim}\Psi_{ni} A=0_{P\times P}
\end{equation}
\begin{equation}
\Psi_{ni}={\sum_{j=1}^{n-1}\Gamma_{ik} \Gamma_{jk}^{\dagger} e_{j}^{T}}-\beta_{P}e_{i}^{T}
\end{equation}
If $P=1$ :
\begin{equation}
\Psi_{N}=\Psi^{+}_{N}
\end{equation}
with the conditions $k\neq j \neq i$ and $\beta_{P}=(n-1)I_{P}$
\end{proposition}
If we take an example of an array consisting of $N=500$ cells and five statistically independent sources with moderate $SNR$ and sufficient number of snapshots $K$, we have $card(\Omega)=\left (\sum_{i=2}^{100}  i \right )=5049$ possibilities $\Omega=\left \{\Psi_{2},\Psi_{3},...,\Psi_{100} \right \}$  where the limit partition is $n=100$. To make the idea more clear, we present in the next section an example for $n=3$.
\section{Elementary example, $n=3$} 
In this example, we suppose that $n=3=\lfloor N/P \rfloor$, the constants become $\beta_{P}=2I_{P}$, $i=\{1,2,3\}$ and the generated base is $\{e_{i}\}_{1\leq i \leq 3}$, the channel matrix is decomposed into the following :
\begin{equation}
A=e_{1} e_{1}^{T}A+e_{2}e_{2}^{T}A+e_{3}e_{3}^{T}A
\end{equation}
\begin{equation*}
=e_{1}A_{1}+e_{2}A_{2}+e_{3}A_{3}
\end{equation*}
The dimensions of the three blocks are $A_{1} \in \mathbb{C}^{P\times P}$, $A_{2} \in \mathbb{C}^{P\times P}$ and $A_{3} \in \mathbb{C}^{N-2P\times P}$ as :
\begin{equation}
A= \left ( \begin{array}{rrr}
A_{1}\\
A_{2}\\
A_{3}\\
\end{array} \right )
\end{equation}
The partial operators are given by :
\begin{equation}
A_{1}=Q_{12}A_{2}, A_{1}=Q_{13}A_{3}, A_{2}=Q_{23}A_{3}, Q_{13}=Q_{12}Q_{23}
\end{equation}
Consequently the spectral matrix is partitioned into the following :
\begin{equation}
\Gamma= \left ( \begin{array}{ccc}
\tilde{\Gamma}_{11}    & \Gamma_{12}     &   \Gamma_{13}  \\
\Gamma_{21}            & \tilde{\Gamma}_{22} & \Gamma_{23}\\
\Gamma_{31}            & \Gamma_{32}      & \tilde{\Gamma}_{33}
\end{array}\right )
\end{equation}
The index $k$ has $n-2=1$ possibility, let us compute the first propagator :
\begin{equation}
\Psi_{31}=-2e_{1}^{T}+\sum_{j=2,k \neq j}^{3}
\Gamma_{1k} \Gamma^{\dagger}_{jk}e_{j}^{T}
\end{equation}
\begin{equation*}
=-2e_{1}^{T}+\Gamma_{13}\Gamma^{\dagger}_{23} e_{2}^{T} +\Gamma_{12}\Gamma^{\dagger}_{32} e_{3}^{T}
\end{equation*}
Similarly we obtain the two left propagators as follows :
\begin{equation}
\Psi_{32}=\Gamma_{23}\Gamma^{\dagger}_{13} e_{1}^{T}-2e_{2}^{T}+\Gamma_{21}\Gamma^{\dagger}_{31} e_{3}^{T}
\end{equation}
\begin{equation}
\Psi_{33}=\Gamma_{32}\Gamma^{\dagger}_{12} e_{1}^{T}+\Gamma_{31}\Gamma^{\dagger}_{21} e_{2}^{T}-2e_{3}^{T}
\end{equation}
The assembled matrix $\Psi_{3}$ is  given by :
\begin{equation}
\Psi_{3}=\left (  \begin{array}{ccc}
-2I_{P}                     & \Gamma_{13}\Gamma^{\dagger}_{23} & \Gamma_{12}\Gamma^{\dagger}_{32} \\
\Gamma_{23}\Gamma^{\dagger}_{13} & -2I_{P}                     & \Gamma_{21}\Gamma^{\dagger}_{31} \\
\Gamma_{32}\Gamma^{\dagger}_{12} & \Gamma_{31}\Gamma^{\dagger}_{21} & -2I_{N-2P}
\end{array}   \right )
\end{equation}
A particular case  $\Psi_{33}$ in this example was presented in [6], where it was stated that the method is comparable to the standard Propagator and standard MUSIC estimators when the SNR is sufficiently high, besides it was also stated that $\Psi_{33}$ can resolve two closely spatial sources if the noise is spatially non uniform. The other operators $\Psi_{31}$ and $\Psi_{32}$ are not mentioned in the literature.

\section{summary}
In this brieve section, we describe the algorithm presented earlier :\\
\begin{table}[!h]
\centering
\caption{ Extended Propagator algorithm.}
\begin{tabular}{|p{15.5cm}|}

\hline
Input : $(\Gamma,P)$.\\

1.Choose $n$ from $2\leq n \leq \lfloor N/P \rfloor$.\\

2.Generate $\{e_{i}\}_{1\leq i \leq n} \in \mathbb{R}^{N\times P}$.\\

3.Choose $p$ from $\{ 1,2,...,n\}$.\\

4.Compute $\Psi_{np}={\sum_{j=1}^{n}\Gamma_{pk} \Gamma_{jk}^{\dagger} e_{j}^{T}}-\beta_{p}e_{p}^{T}$.($k \neq j\neq p$).\\

\hline
\end{tabular}
\end{table}

In the following table we present different cases encountred in real life applications and the cases where the proposed formalism is available :\\ \\
\begin{table}[!h]
\centering
\begin{tabular}{|p{2.5cm}||p{13cm}|}
\hline
$0 < n \leq 1$ &  $\Omega=\emptyset$ , possibility exists for quasi-stationary signals using KHATRI RAO space [15].\\
\hline
$1 <  n < 2 $  &  $\Omega=\emptyset$,  Standard Propagator is available.\\ \hline
$2 \leq n \leq \lfloor N/P \rfloor $ &  $\Omega=\{\Psi_{2},...,\Psi_{n} \}$, $n(n+1)/2-1$ possibilities of extended propagators.\\ \hline
\end{tabular}
\end{table}

The proposed operators are  valid for arbitrary geometry as long as the distance between consecutive sensors is about half the wavelength :
\begin{equation}
|| \vec{r}_{i+1}-\vec{r}_{i} || \leq {\lambda \over 2}
\end{equation}
With $\vec{r}_{i}$ being the position of the $i^{th}$ sensor per given reference.\\
The accuracy of the spectra is measured by comparing theirs variance with the Cramer Rao Bound (CRB). But in terms of localization function, we propose the following definition of the ideal spectrum. The one dimensional ideal spectrum $f(\theta)$ for detecting $P$ sources is Dirac comb :
\begin{equation}
\underset{SNR \to +\infty}{\lim} f(\theta)= \sum_{k=1}^{P} \alpha_{k}\delta(\theta-\theta_{k})
\end{equation}
with $\theta \in \left [ -{\pi \over 2},+{\pi \over 2}\right ]$. In other formulation, the ideal spectrum is the one sided  Fourier transform of superposition of $P$ sinusoidals sampled with frequency $F \geq\pi$ :
\begin{equation}
f(q)=\sum_{k=1}^{P} \Re\{ \alpha_{k} e^{j\omega_{k}q} \}
\end{equation}
With $\omega_{k}=2\pi \theta_{k}$ and $q=\theta^{-1}$, most of the subspace based techniques do not directly provide the information about the magnitudes $\alpha_{k}$, the obtained peaks measure the degree of orthogonality between the signal and noise subspaces. In the next section we present some simulation results to test the performance of several propagators $\Psi_{np}$ comparatively to some high resolution eigen-based methods such as ESPRIT which is briefly described in the following section.
\subsection{Basic Review of ESPRIT algorithm} 
The ESPRIT algorithm ( Estimation of Signal Parameters via Rotational Invariance) [8]-[16] is high resolution and double eigendecomposition approach, it is based on computing the DoAs from the eigenvalues of rotation operator $\Psi$ that relates two identical partitions of the antenna array given that each element in one subarray  has its own twin in the other, in other words, the array is characterized by displacement invariance $d$. The follwoing table summarizes the algorithm :

\begin{table}[!h]
\centering
\caption{ ESPRIT algorithm.}
\begin{tabular}{|p{15.5cm}|}

\hline
Input : $X\in \mathbb{C}^{N\times K}$.\\

1.Create two partitions of array by choosing the number $m$ defined by : $P\leq m \leq N-1$.\\

2.Compute signal subspace $U_{s}$ from $X=U \Sigma V^{+}$ or from $\Gamma=U \Lambda U^{+}$.\\

3.Generate selection matrices $e_{1}=\left [ I_{m} \mid 0_{m\times (N-m)} \right ] \in \mathbb{R}^{m\times N}$, $e_{2}=\left [ 0_{m\times (N-m)} \mid I_{m} \right ]$ and solve : $e_{1}U_{s} \Psi \simeq e_{2} U_{s}$.\\

4.Compute the eigenvalues of $\Psi=T \Phi T^{-1}$.\\

5.Extract the DoAs by $\mu_{i}=Arg\{ \Phi_{i}\}$ and $\theta_{i}=Arcsin({-\lambda \mu_{i}  /2 \pi d } )$ for $1 \leq i \leq P$ \\
                       \\

\hline
\end{tabular}
\end{table}

\section{SIMULATION RESULTS}
We run some computer simulations to verify the performance of some propagators presented in this paper, for this purpose we consider $P=3$ narrowband punctual sources with carrier frequency of $f_{c}=2GHz$ are impinging on an array consisting of $N=18$ omnidirectional sensors uniformely spaced with half the wavelength $d=7cm$ which makes the total length of the array $L_{\lambda}=119cm$ with corresponding Rayleigh angular limit resolution of $\theta_{HPBW}\simeq 6^{\circ}$.\\
The sources are located in the Fraunhofer region where theirs azimuths, measured from broadside, are $\theta_{1}=10^{\circ}$, $\theta_{2}=21^{\circ}$ and $\theta_{3}=45^{\circ}$. The received signals are equipowered and complex identically distributed random processes $s_{i}(t) \sim \mathcal{CN}(0,\sigma_{s,i}=1 watt)$ and the number of snapshots is $K=200$ samples .\\
The number of possibilities is given by the ensemble $\Omega=\{\Psi_{2}, \Psi_{3}, \Psi_{4}, \Psi_{5}, \Psi_{6} \}$ with $Card(\Omega)=20$, due to the diversity of possibilities we only take some operators for performance evaluation.\\
In the first experiment, we fix $SNR=5dB$, the figure 2 represents an average of $L=200$ Monte Carlo runs of the operators, $\Psi_{21}$,$\Psi_{31}$,$\Psi_{32}$ and $\Psi_{41}$.

\begin{figure}[!h]
\centering
\includegraphics[width=9cm,height=7cm]{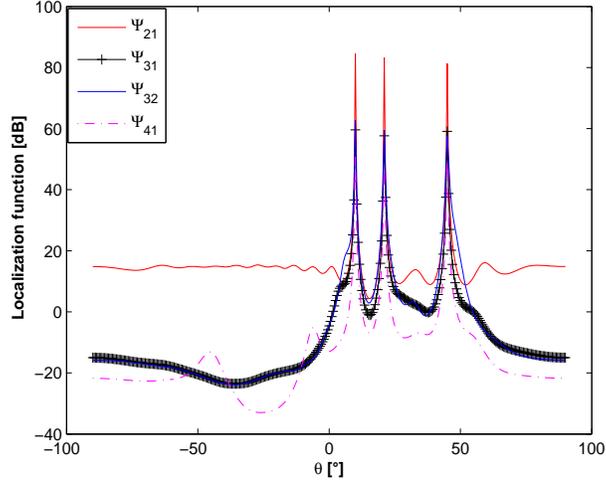}
\caption{Average of $L=200$ Monte Carlo runs of  $ \{ \Psi_{21},\Psi_{31},\Psi_{32},\Psi_{41} \}$ with 
$d=\lambda / 2$, $SNR=5dB$, $K=200$ and $\theta$ = [ $10^{\circ}$, $21^{\circ}$, $45^{\circ}$ ].}
\label{FIG2}
\end{figure}
The peaks are sharp for all the operators and the physical interpretation is easy to make due to lack of sidelobs, however the difference lies in the magnitude of the peaks for each $\Psi_{ij}$.\\
Using the same conditions, we compute the correlation coefficients of $\Psi_{4}$ to explore the relationship between four operators, the following matrix represents the coefficients :
\begin{equation}
C_{\Psi_{4}}=\left ( \begin{array}{cccc}
1.00 & 0.87  & 0.80  & 0.97 \\
0.87  & 1.00 & 0.96 &  0.87\\
0.80 & 0.96 & 1.00 & 0.79 \\
0.97 & 0.87 & 0.79 & 1.00 \\
\end{array}\right )
\end{equation}
We observe that the pairs $(\Psi_{41},\Psi_{44})$ and  $(\Psi_{43},\Psi_{42})$ are highly correlated. The figure 3 represents the Root Mean Square Error (RMSE) of $\Psi_{4}$ comparatively to the standard ESPRIT method [8].
\begin{figure}[!h]
\centering
\includegraphics[width=9cm,height=7cm]{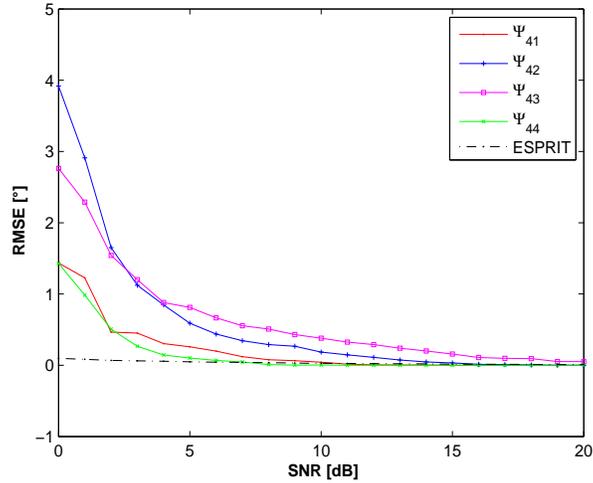}
\caption{Average of $L=200$ Monte Carlo runs of  $\Psi_{4}$ and ESPRIT for each value of $SNR$, 
$d=\lambda / 2$, $K=200$ and $\theta$ = [ $10^{\circ}$, $21^{\circ}$, $45^{\circ}$ ].}
\label{FIG3}
\end{figure}
When SNR is about $20dB$, all the spectra are equivalent, in the other hand we realize that $(\Psi_{41},\Psi_{44})$ vary with the same rate, which is also the same case for $(\Psi_{43},\Psi_{42})$, this remark is confirmed by the matrix $C_{\Psi_{4}}$.\\
In the last experiment, we compare the spectrum of $\Psi_{61}$ with that MUSIC [7] and G-MUSIC [2], the last method outperforms the two other while the function of   $\Psi_{61}$ presents several fluctuations as minor sidelobs.

\begin{figure}[!h]
\centering
\includegraphics[width=9cm,height=7cm]{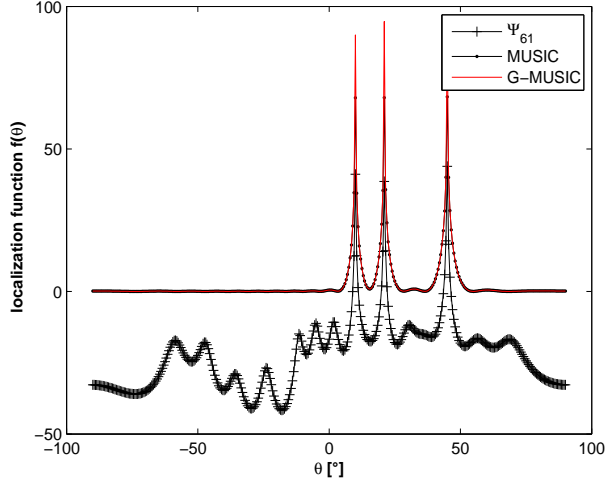}
\caption{Average of $L=200$ Monte Carlo runs of  three spectra with $SNR=10dB$, $d=\lambda / 2$, $K=200$ and $\theta$ = [ $10^{\circ}$, $21^{\circ}$, $45^{\circ}$ ].}
\label{FIG4}
\end{figure}

Some of the immediate enhancements that can be added to the present paper is the orthonormalization [4] of any propagator $\Psi_{np}$ which will add more resolution power in source detection. The other proposition lies in using the real valued transformation [17] of the spectral matrix $\Gamma$, this will reduce the computational complexity and increase the probability of locating the coherent sources.\\
In the other hand, the drawback of the proposed formalism resides in noise condition, the $SNR$ must be moderate and the number of radiating sources $P$ must be known, hence a non eigen-based spectral method to determine P will complete the extended propagator algorithm.
\section{Conclusion}
In this paper, we have introduced a proposition which states that when the ratio $n$ of the number of antenna sensors $N$ to the number of narrowband non coherent sources $P$ is superior or equal to 2, and the $SNR$ is sufficiently high, it is possible to generate ${n(n+1)/2}-1$ distinct versions of high resolution propagator operators to detect the angular positions of the radiating sources with no constraint on the geometry of the array.\\
Original formalism is presented, inspired from semiclassical approach of statistical mechanics, to give the limit of the channel matrix partition, from which diversity of extended propagators is derived, this new approach is suitable for large array. In Monte Carlo simulation results, we have presented elementary examples to test the performance of the extended propagators of orders $n=3$, $n=4$ and $n=6$ which proved theirs resolving powers in high SNR conditions.
\section{Appendix}
In this section we present parts of computer programs implemented in simulations, using MATLAB software
 
\begin{itemize}
\item {\bf Operators $\Psi_{31}$ and $\Psi_{32}$} :\\
  N=18;\\
  P=3;\\
  R12=R(1:P,P+1:N);\\
  R13=R(1:P,2*P+1:N);\\
  R23=R(P+1:2*P,2*P+1:N);\\
  R12=R(1:P,P+1:2*P);\\
  R21=R12';\\
  R32=R23';\\
  R31=R13';\\
  Q31=[-2*eye(P) R13*pinv(R23) R12*pinv(R32)];\\
  Q32=[R23*pinv(R13) -2*eye(P) R21*pinv(R31)];\\

\item {\bf Operator $\Psi_{41}$} :\\
 R13=R(1:P,2*P+1:3*P);\\
 R23=R(P+1:2*P,2*P+1:3*P); \\
 R34=R(2*P+1:3*P,3*P+1:N); \\
 R14=R(1:P,3*P+1:N); \\
 R42=R(3*P+1:N,P+1:2*P); \\
 R12=R(1:P,P+1:2*P); \\
 Q41=[-3*eye(P) R13*pinv(R23) R14*pinv(R34) R12*pinv(R42)]; \\
\item {\bf Operator $\Psi_{61}$ }:\\
 R12=R(1:P,P+1:2*P);\\
 R13=R(1:P,2*P+1:3*P);\\
 R23=R(P+1:2*P,2*P+1:3*P);\\
 R14=R(1:P,3*P+1:4*P);\\
 R34=R(2*P+1:3*P,3*P+1:4*P);\\
 R15=R(1:P,4*P+1:5*P);\\
 R45=R(3*P+1:4*P,4*P+1:5*P);\\
 R16=R(1:P,5*P+1:N);\\
 R56=R(4*P+1:5*P,5*P+1:N);\\
 R63=R(5*P+1:N,2*P+1:3*P);\\
 Q61=[-5*eye(P) R13*pinv(R23) R14*pinv(R34) R15*pinv(R45) R16*pinv(R56) R13*pinv(R63)];\\
\end{itemize}

\end{document}